\documentclass{article}

\usepackage[final]{nips_2017}

\usepackage{amsmath}
\usepackage[utf8]{inputenc} % allow utf-8 input
\usepackage[T1]{fontenc}    % use 8-bit T1 fonts
\usepackage{hyperref}       % hyperlinks
\usepackage{url}            % simple URL typesetting
\usepackage{booktabs}       % professional-quality tables
\usepackage{amsfonts}       % blackboard math symbols
\usepackage{nicefrac}       % compact symbols for 1/2, etc.
\usepackage{microtype}      % microtypography

\usepackage{graphicx}

\title{Does Phase Matter For Monaural Source Separation?}

\author{
  Mohit Dubey\\
  Oberlin College and Conservatory\\
  \texttt{mdubey@oberlin.edu} \\ 
  \And
 Garrett Kenyon \\
  Los Alamos National Laboratory\\
  \texttt{gkenyon@lanl.gov} \\
  \AND  
  Nils Carlson \\
  New Mexico Institute of Mining and Technology\\
  \texttt{nils.carlson@student.nmt.edu} \\
   \And  
  Austin Thresher \\
  New Mexico Consortium \\
  \texttt{athreshe@gmail.com} \\
}

\begin{document}
% \nipsfinalcopy is no longer used

\maketitle

\begin{abstract}
The "cocktail party" problem of fully separating multiple sources from a single channel audio waveform remains unsolved. Current biological understanding of neural encoding suggests that phase information is preserved and utilized at every stage of the auditory pathway. However, current computational approaches primarily discard phase information in order to mask amplitude spectrograms of sound. In this paper, we seek to address whether preserving phase information in spectral representations of sound provides better results in monaural separation of vocals from a musical track by using a neurally plausible sparse generative model. Our results demonstrate that preserving phase information reduces artifacts in the separated tracks, as quantified by the signal to artifact ratio (GSAR). Furthermore, our proposed method achieves state-of-the-art performance for source separation, as quantified by a mean signal to interference ratio (GSIR) of 19.46.
\end{abstract}

\section{Introduction}

The "cocktail party problem" - isolating a sound-stream of interest from a complex or noisy mixture - remains one of the most important unsolved problems in signal processing. While it is well understood that the human (mammalian) auditory system is an expert at solving this problem, even without binaural or visual cues, it is still unclear how a similar solution can be implemented computationally. In biological systems, it is understood that bottom-up grouping of similar sounds (harmonic stacks), top-down attention towards sounds of interest ("voice" vs. "guitar"), as well as amplitude fluctuations of sounds in similar frequency bands (vibrato/tremolo) are all useful processes for blind monaural source segregation [1]. The extreme degree of overcompleteness of the auditory cortex when compared to the cochlea implies that sparse encoding of auditory information is another key part of the neural solution [2]. \\

 Current understanding of neural auditory encoding consists of two distinct theories: "place theory" and "volley theory". In  place theory, different frequencies of sound are encoded in different physical locations, as seen in the spectral tuning of the cochlea and the resulting tonotopic map that propagates all the way up to the primary auditory cortex (A1) [3]. In volley theory, the frequency and phase of a sound wave are encoded by neuronal spikes that coincide with a regular point on that wave, a process known as "phase-locking" [4]. Research on the mammalian auditory pathway has shown that "phase locking" in the auditory nerve can operate at frequencies up to 4 - 6 kHz (depending on the animal) and is preserved throughout the auditory pathway to A1 at frequencies up to \textasciitilde200 Hz, allowing for phase-locking to the fundamental frequencies of the human voice and many musical instruments [5]. The preservation of phase information throughout the auditory path is useful in binaural source separation as well as speech processing, as A1 cortical neurons have been shown to phase-lock to a sequence of phonemes [5] [6]. A1 cortical neurons in cats have also been shown to convey more abstract auditory information than structured spectro-temporal patterns, implying that A1 receptive fields are complex and well tuned for monaural source separation [7]. \\

In recent years, multifarious computational approaches have tried to solve the blind monaural cocktail party problem including robust PCA [8], Recurrent Neural Networks [9], Convolutional Denoising Autoencoders [10], and Non-Negative Matrix Factorization [11]. A large majority of these implementations separate sources by using some form of "masking" of amplitude power spectra, discarding all phase information from the original sound during training [12] and only preserving time-frequency bins that (up to some threshold) contain the source of interest. A recent work utilizing a Fully Complex Deep Neural Network (FDCNN) has shown that preserving both the real and imaginary parts of the Fourier representations of musical sound during training provides better results compared to a non-complex DNN, especially when a sparseness penalty is also employed [13]. While the current literature has provided insight into discriminative modeling for monaural source separation, it is still not clear whether preserving phase information (as the brain does) assists in learning the underlying generators of distinct sound sources in a sparse encoding. \\

	The idea of sparse approximation is based on the principle that the brain uses as few active neurons as possible to represent a signal as accurately as possible. It has been shown that by learning sparse representations of static images or movies, networks converge on features similar to the receptive fields of simple cells in the primate primary visual cortex [14] [15] [16]. Analogously, one might expect to learn features similar to the spectrotemporal receptive fields present in the primary auditory cortex when learning on phase-rich spectrotemporal representations of auditory input. This analogy has been explored using the locally competitive algorithm (LCA) on time-dependent power spectral and cochlear representations of speech and music with the resulting features exhibiting many of the properties of physiological receptive fields [17] including phase information [18]. \\

	 In this paper we demonstrate that preserving phase information in learning a sparse generative model of musical data results in better separation of vocals from a musical track. We trained three sparse representations on musical audio using a convolutional locally competitive sparse coding algorithm using PetaVision (https://github.com/PetaVision): one preserving phase, two discarding phase. We judged the relative success of the results on the source separation task based on the SDR (Signal-to-Distortion), SIR (Signal-to-Interference), and SAR (Signal-to-Artifact) ratios [19].

\section{Methods}

Our models were trained using the MIR-1K dataset [20] which consists of 1000 song clips ranging from 4 to 16 seconds sampled at 16 kHZ. The clips are taken from recordings of male and female amateur singers singing along to 110 Chinese karaoke tracks. The vocal and non-vocal tracks are stored in separate stereo channels, making the dataset ideal for monaural source separation studies. For this study, we shortened all of the clips to 2 seconds for computational ease. The dataset was divided into 950 training and 50 testing examples which were both randomly sorted, such that multiple male and female singers were present throughout. \\

For the resultant training and testing sets, single channel mixtures were created with the vocals and non-vocals mixed equally and sampled at 16 kHZ. Spectral representations of the vocals, non-vocals, and mixtures for both train and test sets were generated using a 512 point short time Fourier transform (STFT) with 50 percent overlap and a Hanning Window, resulting in 256 independent frequency bins up to 8 kHz and 128 time bins. Phase information (both real and imaginary parts) were stored in one case, while only the modulus amplitude was stored in the other (no phase information). Logically, the phase rich representations contain twice the amount of information as the representation without phase. To avoid this bias, a third pair of train and test sets was also generated without phase information using the same short time Fourier transform (STFT) but with twice the number of independent frequency bins (512). \\

All three datasets of vocal and non-vocal mixtures were then trained identically using the Locally Competitive Algorithm (LCA), which functions by minimizing the error between input and reconstruction while also minimizing the activity of elements used in the representation. This is expressed mathematically as the minimization of an energy function, \\

\begin{equation}
E(\overrightarrow{I}, \boldsymbol{\phi}, \overrightarrow{a}) = \min\limits_{ \{\overrightarrow{a}, \, \phi \} } \left[  \, \frac{1}{2}  ||  \overrightarrow{I} - \boldsymbol{\phi} \overrightarrow{a} ||^2	+	\lambda || \overrightarrow{a} ||_1 \right]
\label{eq:SC}
\end{equation}

where $\overrightarrow{I}$ represents the input, $\overrightarrow{a}$ represents activation values for the basis vectors $\phi$, and $\lambda$ represents the threshold or "sparsity penalty". Approximate minimization of the energy function for a given two second "sound image" is accomplished by numerically solving the ordinary differential equation that results from taking the partial derivative of the energy with respect to activation coefficients, while a minimum of the energy function  with respect to the basis vectors is accomplished by iterative stochastic gradient descent using a local Hebbian rule with momentum. In order to obtain a more neurally plausible network, our implementation deploys horizontal competition between elements in the hidden layer, by inserting membrane potentials $u$ such that $\overrightarrow{a} = T(u)$, where $T(u)$ is a soft-threshold function, and $\frac{du}{dt}$ is the derivative of the energy function with respect to the activation coefficients (as in Rozell et al. [16]) . \\

To locate the optimum threshold for our datasets, we first trained our network for two epochs on the mixed training set (phase only) at multiple threshold values. We then added Gaussian noise to the training set and used the resulting sparse codes to denoise the data. Through this method, we found that the optimum threshold for our network was around .625, resulting in average sparsity of approximately 2.8 percent (see Figure 1). \\

\begin{figure}[h]
 \centerline{\includegraphics[scale=.9]{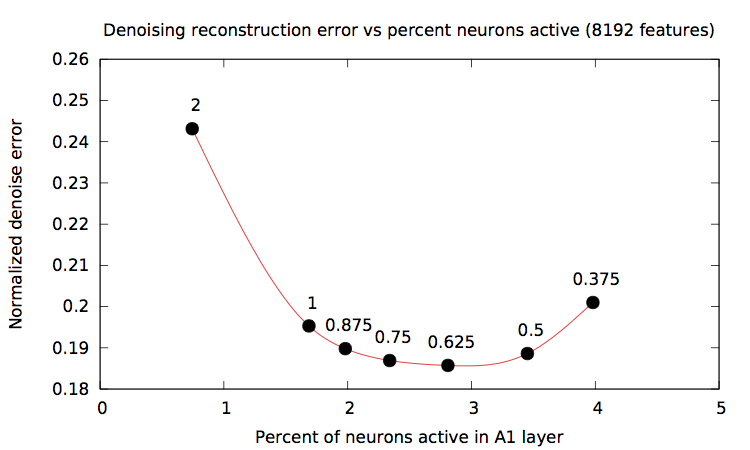}}
  \caption{Normalized denoising error for various thresholds (labeled) plotted by average sparsity.}
\end{figure}

Each network was trained for four epochs on the respective training set (Phase, No Phase, No Phase x2) using a display period of 1000 timesteps, a dictionary size of 8192 features (2x overcomplete) and stride of 2. Each feature represented the entire frequency dimension and 128 ms of sound. The resultant dictionaries were used to write out sparse codes for  the training and testing sets of the three mixtures. \\

Once sparse codes were generated for the three datasets, a linear classifier was implemented in Petavision to separate out vocals and non-vocals from the mixtures. The classifier was initialized with the features trained during the sparse coding procedure and allowed to adapt for 40 epochs through the training set. All other parameters including the threshold and learning rate were held constant from the sparse coding procedure. The resulting features were used to separate vocals and non-vocals on the three testing sets and the resulting stems were used to calculate the GSIR, GSAR, and GNSDR, as shown in [8]. A fourth set of stems were generated by running the phase-rich results through the classifier again to denoise the separated vocal and non-vocal tracks, providing the best possible results for our blind monaural source separation technique (Denoised).

\section{Results}

Performance results for each of the datasets are provided in Table 1.  GSIR measures the remaining interference between the sources, GSAR indicates the artifacts caused by the separation algorithm and GNSDR measures how distorted the separated sources are and is considered the overall performance evaluation for source separation. While it is clear that GNSDR and GSIR show no general improvements with phase information included, 	the GSAR is noticeably higher for the phase results. This implies that including phase information introduces fewer artifacts during the source separation process, even when more frequency bins are used to account for the loss of information. Furthermore, our mean SIR value of 19.46 for the denoised audio indicates very high separation quality that is state of the art compared to other published methods [8][12][13].

\begin{figure}[h]
 \centerline{\includegraphics[scale=0.5]{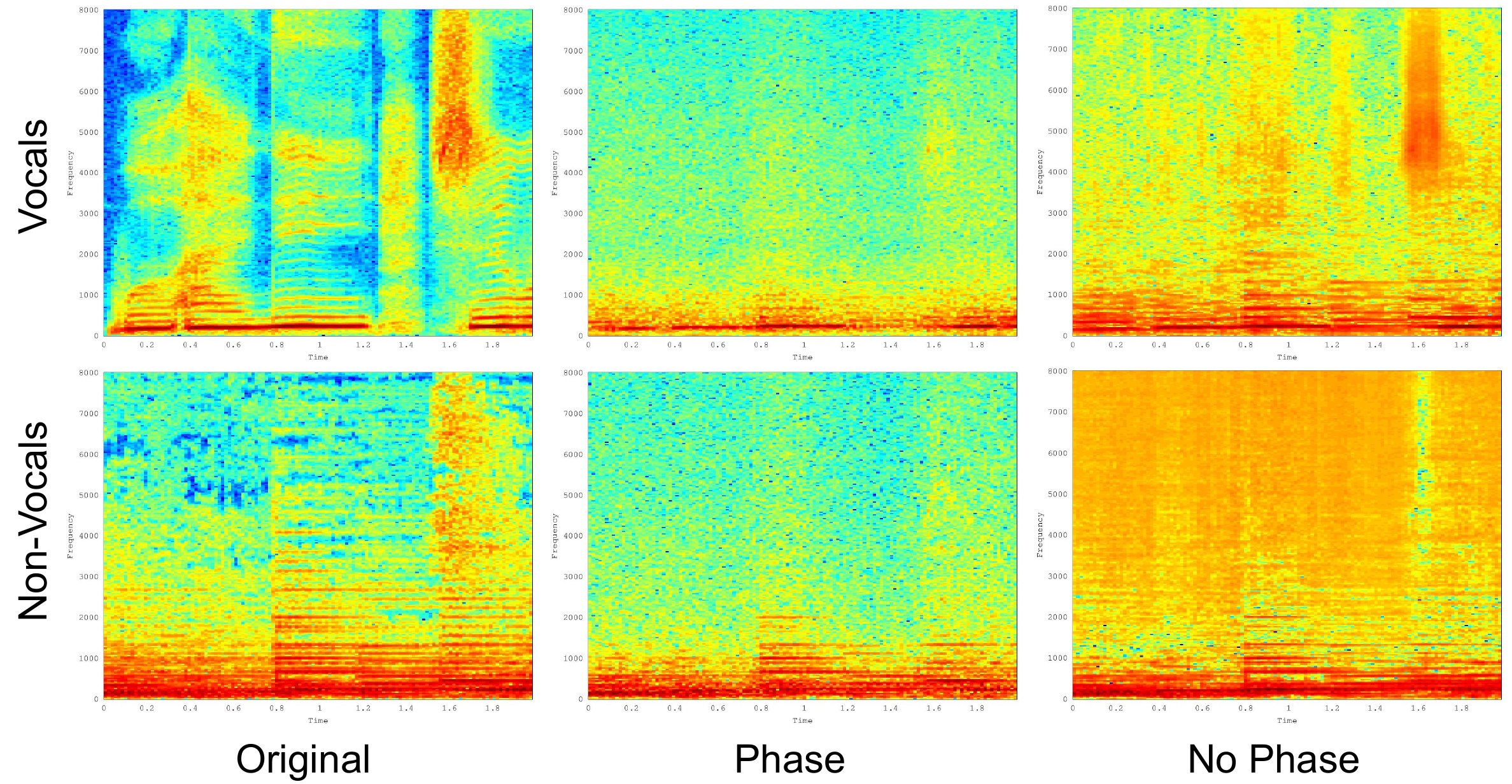}}
  \caption{Spectrograms of non-vocals and vocals for phase and no-phase of a sample from the test set. While phase reconstructs less high-frequencies, no-phase adds energy in spurious places (artifacts). }
\end{figure}

\begin{table}[t]
  \caption{Results on Test Set}
  \label{sample-table}
  \centering
  \begin{tabular}{llllll}
    \toprule
    \cmidrule{1-2}
    Run     & GSIR     & GSAR & GNSDR  \\
    \midrule
    Phase 		& $12.12 \pm 9.12$   & $-3.51 \pm 5.92$   &  $1.49 \pm 3.02$  \\
    No Phase     & $12.76 \pm 12.07$   & $-9.72 \pm 4.36$   &  $1.35 \pm 4.01$      \\
    No Phase x 2     & $14.84 \pm 13.26$   & $-10.95 \pm 4.50$   &  $1.86 \pm 3.88$  \\
    Denoised 	& $19.46 \pm 8.61$   & $2.88 \pm 5.54$   &  $5.21 \pm 2.77$
 
  \end{tabular}
\end{table}

\section{Conclusion}

Using a generative sparse coding model we have demonstrated that preserving phase information results in fewer artifacts on musical source separation tasks. In addition, we have introduced a new method for separating musical sources that achieves state of the art separation quality. In the future, we plan to run source separation experiments on other auditory data (speech, instrumental music, etc.) as well as explore the benefits of hierarchical sparse coding in source separation. 

\subsubsection*{Acknowledgments}

This work was funded by NSF and the DARPA UPSIDE program. 

\section*{References}

\small

[1] McDermott, J. H. (2009). "The cocktail party problem." Current Biology, 19(22), 1024-1027. Retrieved June 8, 2017.

[2] Chechik, Gal, Michael J. Anderson, Omer Bar-Yosef, Eric D. Young, Naftali Tishby, and Israel Nelken. "Reduction of Information Redundancy in the Ascending Auditory Pathway." Neuron 51 (2006): 359-68. Print. 
	
[3] Oxenham, A. J. (2013). "Revisiting place and temporal theories of pitch." Acoustic Science and Technology,34(6), 388-396.			
			
[4] Joris, P. X. (2004). "Neural Processing of Amplitude-Modulated Sounds." Physiological Reviews, 84(2), 541-577. doi:10.1152/physrev.00029.2003

[5] Nourski, K. V., and Brugge, J. F. (2011). "Representation of temporal sound features in the human auditory cortex." Reviews in the Neurosciences, 22(2). doi:10.1515/rns.2011.016

[6] Wang, R., Perreau-Guimaraes, M., Carvalhaes, C., and Suppes, P. (2012). "Using phase to recognize English phonemes and their distinctive features in the brain." Proceedings of the National Academy of Sciences of the United States of America, 109(50), 20685-20690. http://doi.org/10.1073/pnas.1217500109

[7] Chechik, G., and Nelken, I. (2012). "Auditory abstraction from spectro-temporal features to coding auditory entities." Proceedings of the National Academy of Sciences of the United States of America, 109(46), 18968-18973. http://doi.org/10.1073/pnas.1111242109

[8] P. S. Huang, S. D. Chen, P. Smaragdis and M. Hasegawa-Johnson, "Singing-voice separation from monaural recordings using robust principal component analysis," 2012 IEEE International Conference on Acoustics, Speech and Signal Processing (ICASSP), Kyoto, 2012, pp. 57-60. doi: 10.1109/ICASSP.2012.6287816

[9] P. S. Huang, M. Kim, M. Hasegawa-Johnson and P. Smaragdis, "Joint Optimization of Masks and Deep Recurrent Neural Networks for Monaural Source Separation," in IEEE/ACM Transactions on Audio, Speech, and Language Processing, vol. 23, no. 12, pp. 2136-2147, Dec. 2015. Doi: 10.1109/TASLP.2015.2468583

[10] Grais, E. M., and Plumbley, M. D. (2017). "Single Channel Audio Source Separation using Convolutional Denoising Autoencoders." ArXiv. Retrieved from https://arxiv.org/abs/1703.08019.

[11] T. Barker and T. Virtanen, "Blind Separation of Audio Mixtures Through Nonnegative Tensor Factorization of Modulation Spectrograms," in IEEE/ACM Transactions on Audio, Speech, and Language Processing, vol. 24, no. 12, pp. 2377-2389, Dec. 2016.

[12] Simpson, A. J., Roma, G., and Plumbley, M. D. (2015). "Deep Karaoke: Extracting Vocals from Musical Mixtures Using a Convolutional Deep Neural Network." ArXiv. doi:	arXiv:1504.04658

[13] Y. S. Lee, C. Y. Wang, S. F. Wang, J. C. Wang and C. H. Wu, "Fully complex deep neural network for phase-incorporating monaural source separation," 2017 IEEE International Conference on Acoustics, Speech and Signal Processing (ICASSP), New Orleans, LA, 2017, pp. 281-285. 		
					
[14] B. Olshausen and D. Field, "Emergence of simple-cell receptive field properties by learning a sparse code for natural images," in Nature, vol. 681, p. 607-609, 1996.
					
[15] B. Olshausen, "Highly overcomplete sparse coding." SPIE 2013
					
[16] C. Rozell, et al., "Sparse coding via thresholding and local competition in neural circuits," in Neural Computation, 2008. 
				
[17] Carlson NL, Ming VL, DeWeese MR (2012), "Sparse Codes for Speech Predict Spectrotemporal Receptive Fields in the Inferior Colliculus." PLoS Comput Biol 8(7): e1002594. doi:10.1371/journal.pcbi.1002594 

[18] Dubey, M. L., Shultz, P. F., and Kenyon, G. T. (2016, March). Learning phase-rich features from streaming auditory images. In Image Analysis and Interpretation (SSIAI), 2016 IEEE Southwest Symposium on (pp. 73-76). IEEE.				

[19] E. Vincent, R. Gribonval and C. Fevotte, "Performance measurement in blind audio source separation," in IEEE Transactions on Audio, Speech, and Language Processing, vol. 14, no. 4, pp. 1462-1469, July 2006. doi: 10.1109/TSA.2005.858005			
		
[20] Chao-Ling Hsu and Jyh-Shing Roger Jang, "On the Improvement of Singing Voice Separation for Monaural Recordings Using the MIR-1K Dataset," IEEE Trans. Audio, Speech, and Language Processing,  volume 18, issue 2, p.p 310-319, 2010.	
\end{document}